Average Extinction Values for Each Observing Run

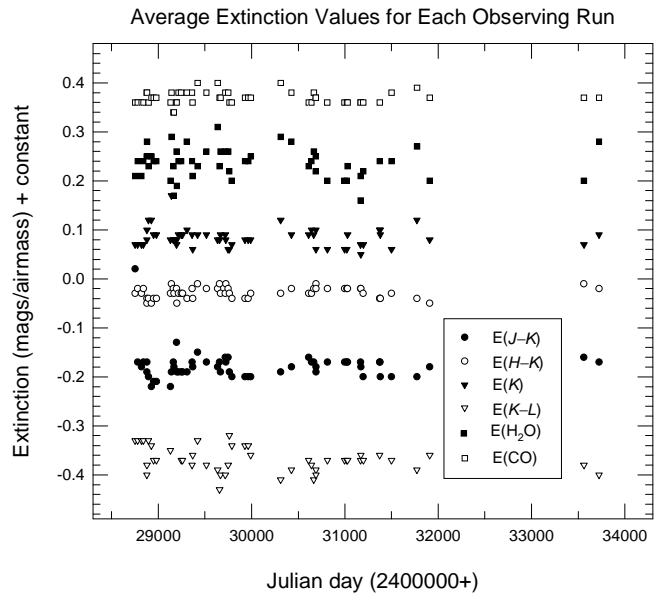

Daily Sensitivities

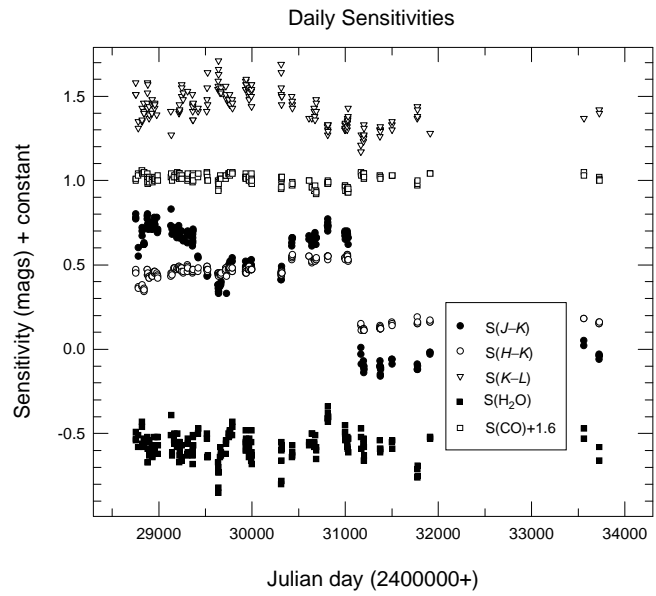

Frogel, IR Extinction, Figures 1a,b

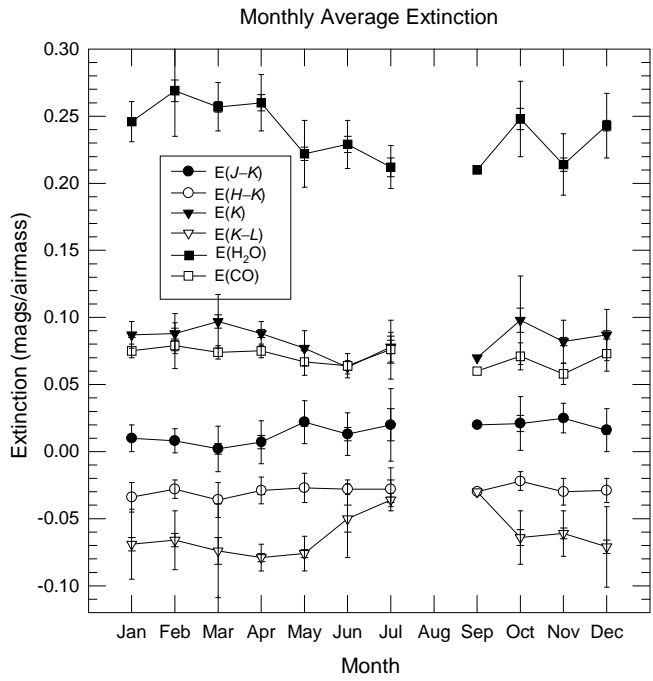

Monthly Average Extinction

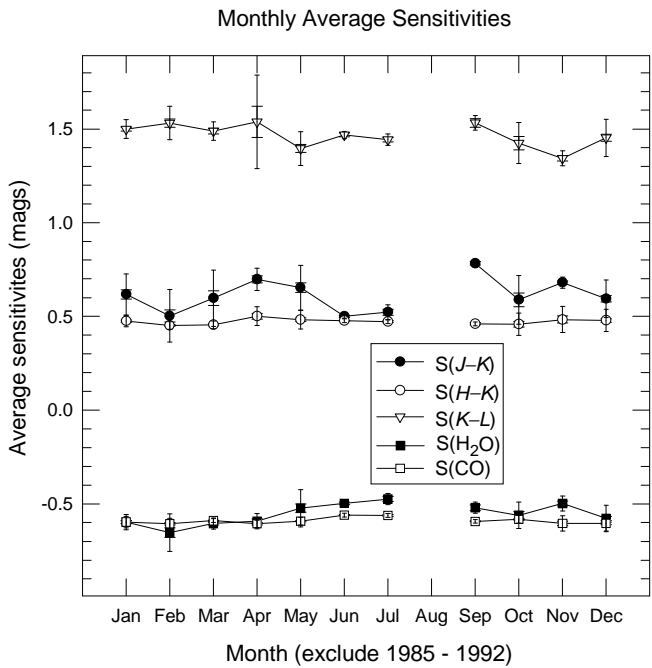

Monthly Average Sensitivities

Frogel, IR Extinctions, Figures 2a,b

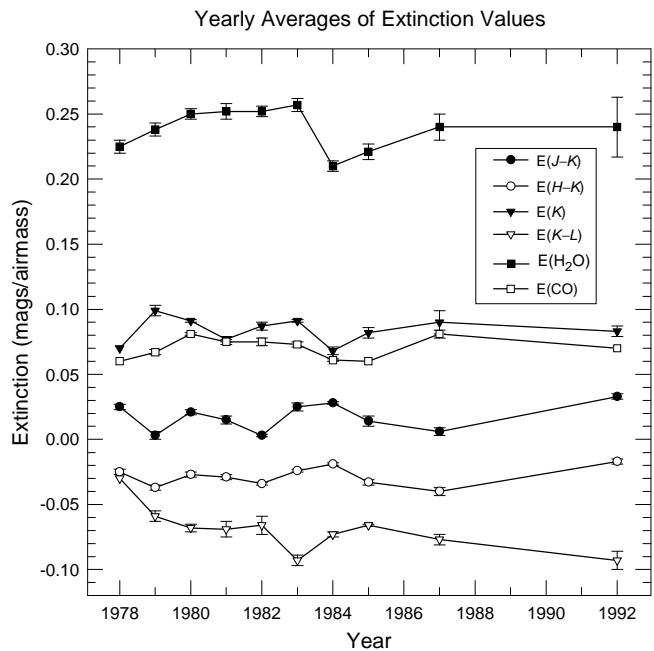

**Yearly Averages of Extinction Values**

Legend:
- E(J−K)
- E(H−K)
- E(K)
- E(K−L)
- E(H$_2$O)
- E(CO)

Axes: Extinction (mags/airmass) vs Year

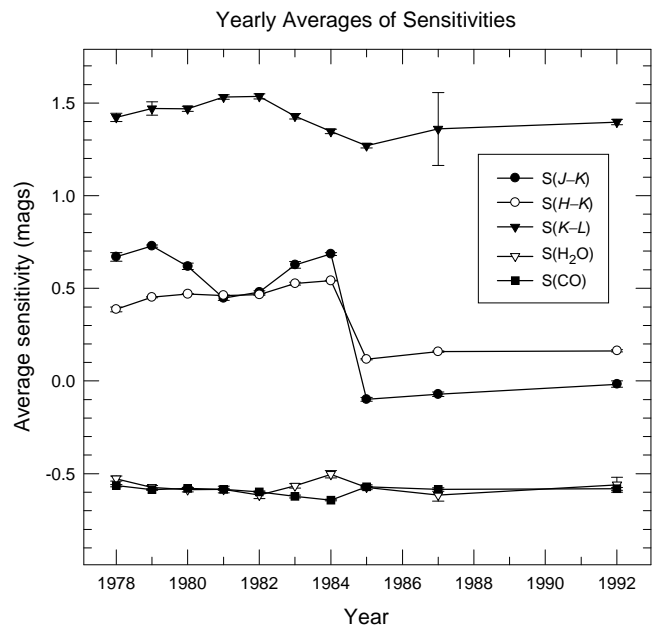

**Yearly Averages of Sensitivities**

Legend:
- S(J−K)
- S(H−K)
- S(K−L)
- S(H$_2$O)
- S(CO)

Axes: Average sensitivity (mags) vs Year

Frogel, IR Extinctions, Figures 3a,b

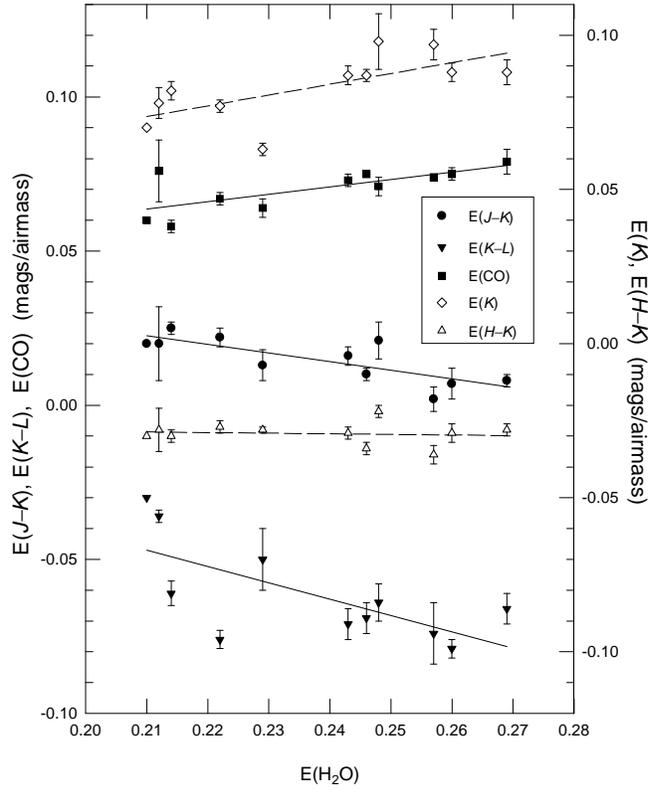

Monthly Averages for Extinction Coefficients

Yearly averages: extinction vs. sensitivity

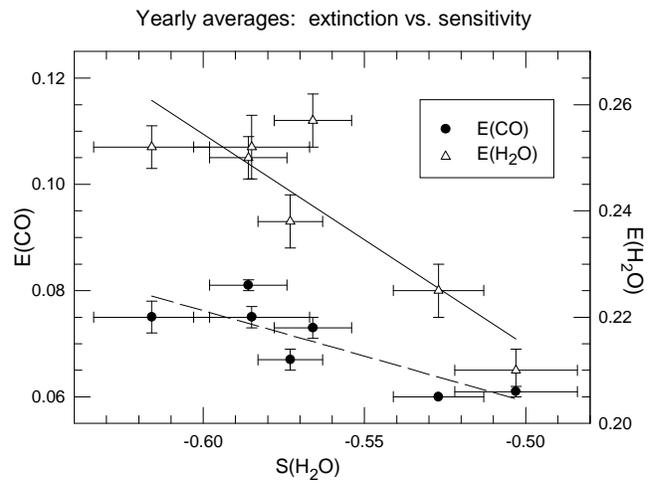

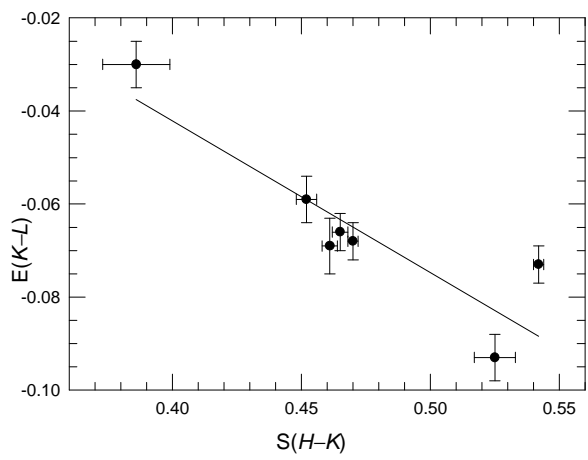

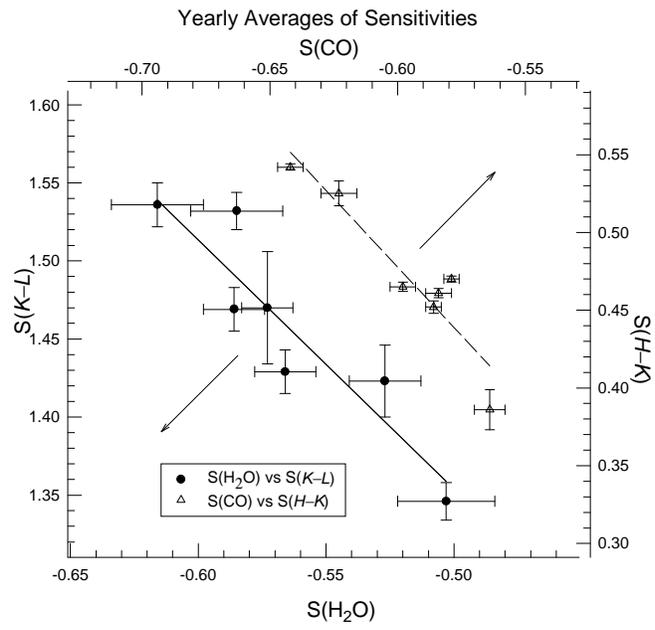

Yearly Averages of Sensitivities

Correlations with ENSO Indices

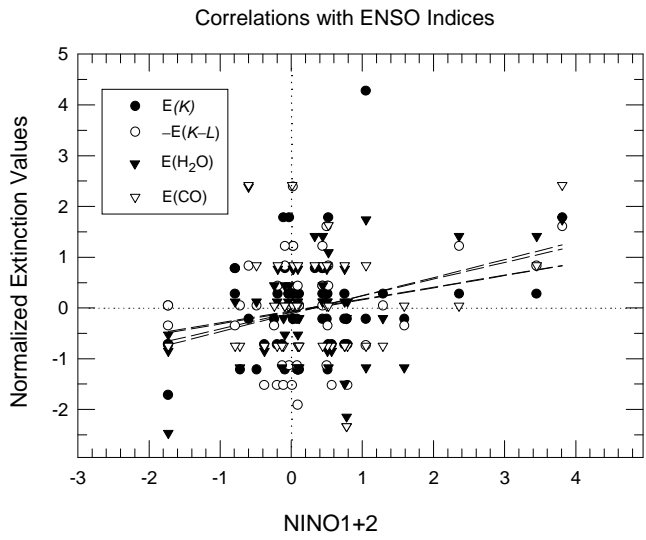

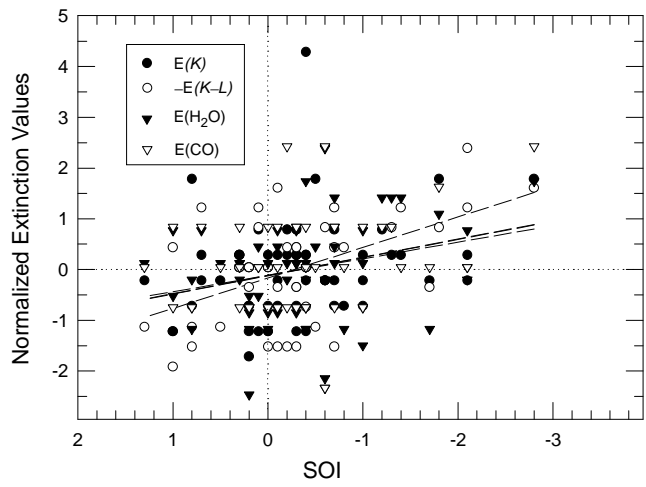

# A HISTORY OF INFRARED EXTINCTION AT CTIO, CHILE, AND A POSSIBLE CONNECTION WITH THE EL NIÑO PHENOMENON

*Jay A. Frogel[1]*

Department of Astronomy, The Ohio State University

174 West 18th Avenue, Columbus, Ohio  43210

Electronic mail:  frogel@galileo.mps.ohio-state.edu

## ABSTRACT

Extinction coefficients and sensitivity values in the *JHKL* band-passes measured on nearly 200 nights of observing between 1978 and 1992 on the 1.5-meter and the Blanco 4-meter telescopes at Cerro Tololo Inter-American Observatory are presented and discussed. Analysis of these data shows the following: There are seasonal variations in both the extinction coefficients and sensitivity values that are qualitatively consistent with expected variations in the amount of $H_2O$ in the atmosphere - relatively high in the summer months, lower in the winter months. The linear correlation coefficients between most of these quantities are statistically significant. The yearly mean values of these quantities also show significant variability of a few hundredths of a magnitude. The correlations between these yearly means are again consistent with variations in the $H_2O$ content of the atmosphere. At least some of these longer term variations are closely correlated with quantitative measures of the strength of the atmospheric and oceanic El Niño/Southern Oscillation phenomenon.

# 1. INTRODUCTION

Unlike atmospheric extinction in optical pass-bands, which is due primarily to molecular diffusion and to scattering by aerosols, both of which are smooth functions of wavelength (*e.g.* Burki *et al.* 1995; Lockwood & Thompson 1986), extinction in the near infrared (near-IR) is due almost entirely to molecular bands of $H_2O$ (primarily) and CO (secondarily). Many of the lines that make up these bands are saturated even under typical conditions; consequently near-IR extinction can behave in a complex manner (see the theoretical study by Manduca & Bell 1979). Because of their origin, temporal variations in the extinction coefficients may be coupled with variations in the "sensitivity" in a filter band-pass, a quantity usually referred to as the "photometric zero-point" in optical photometry.

Between 1978 and 1992 I observed in the near-IR on ~200 photometric nights on the 1.5-m and the Blanco 4-m reflectors at CTIO with the single channel D3 InSb system. These observations provide a uniform, long-term data base of extinction coefficients and sensitivity values which can be investigated for seasonal and secular variations. The effect on atmospheric transparency of the El Niño/Southern Oscillation (ENSO) phenomenon, which has a strong influence on global weather (Cane 1986), can be investigated; so too can effects from volcanic eruptions which are known to reduce atmospheric transmission in the visual (Burki et al. 1995; Lockwood & Thompson 1986; Thompson & Lockwood 1996). Sections 2 and 3 describe the data and the calculation of monthly and yearly averages for the extinction and sensitivity values. Section 3 also examines differences between the two telescopes and other possible instrumental effects. Section 4 addresses the main topic of the paper - seasonal and secular variations in the data. Section 5 gives a brief summary and some conclusions.

# 2. MEASUREMENT OF EXTINCTION COEFFICIENTS AND SENSITIVITIES

The data for this paper were obtained during ~200 nights of observing at CTIO with a single channel near-IR photometer, D3, and with f/30 chopping secondaries. Two thirds of the nights were on the 4-meter Blanco reflector; the remainder on the 1.5-meter. Most of the observations were made from 1978 through 1985 with only 16 nights of observing from 1986 through 1992. Less than 10% of all of the data were obtained in June through September.

The observations were made in the *JHKL* ($\lambda_{eff}$ = 1.25, 1.65, 2.2, & 3.5μm, respectively) band-passes and with pairs of filters that measure stellar absorption due to CO and $H_2O$ in the *K* band-pass (Aaronson et al. 1978; Frogel et al. 1978; Elias et al. 1982). Transmission curves of filters similar to those used for these observations are in Fig. 1. of Manduca & Bell (1979). The 2.00 μm filter, which measures stellar $H_2O$, is the filter most influenced by changes in terrestrial atmospheric transmission. Ferriso et al. (1966, Fig. 2) compares $H_2O$ absorption coefficients at 300 K and 3000 K, temperatures appropriate for the Earth and stellar atmospheres respectively.

Usually, a dozen or more CIT/CTIO standards (Elias et al. 1982) were observed on each night. Instrumental sensitivities, denoted by "S", were determined from these data. The S value for a filter is the average of true minus instrumental magnitude for all standards observed on a night normalized to the same integration time and to one airmass, i.e. overhead. A more positive value of S means greater transmission (atmosphere plus instrument) or brighter instrumental magnitude. Aside from telescope size, S values will vary with time due to seeing, atmospheric transparency, cleanliness of the optics, changes in gain of the electronics, etc. Except for the *K* band, the sensitivities are given for the colors, i.e., the *difference* in sensitivity between two bands, which should be independent of most of these first order effects.

The extinction value for a filter, in units of magnitudes per airmass, measures the attenuation of starlight due to atmospheric opacity. The airmass of an observation has the usual definition, viz., secant $z$, where $z$ is distance from the zenith in degrees. The extinction in a color, *e.g.* E(*J*−*K*), is the simple *difference* in the coefficients for the two filters. Since the magnitudes and colors of the standards are known to an accuracy ≤1%, extinction coefficients were determined by observing, several times during a night, two standards in rapid succession whose airmass differed by ~0.7 or greater, thus eliminating effects of time variable transmission. If the night to night scatter in the coefficients during an observing run, defined as two or more consecutive nights on one telescope, was comparable to or less than those for a single night with multiple determinations of the coefficients – a situation that obtained for nearly all runs – the nightly values were averaged and used for the entire run. Typically, the scatter in the extinction coefficients over an observing run was only slightly greater than expected from uncertainties in the observations of the standards themselves, ≤0.02 mags for all magnitudes and colors. In almost all cases, the need for two sets of extinction corrections during a run was associated with a significant change in the weather, *e.g.* a large shift in night-time temperature or humidity accompanying passage of a frontal system.

## 3. PRESENTATION OF THE DATA

Figures 1a and 1b display the average values of the extinction coefficients for each observing run and the color sensitivities for *every* photometric night of every run, respectively.

### 3.1 Is There A Difference Between Telescopes?

There are no significant differences in the extinction coefficients for the two telescopes. Table 1 gives the global means and standard deviations of these values. All of these differences are ≤0.01 mag. Even those that appear greater than the uncertainties in the means, $\sigma/\sqrt{N}$, are still considerably less than the observed seasonal variations; and there was a seasonal bias in which telescope was used. Combined values are given in the two right-most columns of Table 1 and will be used in the remainder of this paper.

Table 2 presents the sensitivities for the two telescopes. Only pre-1985 data were considered since, as is evident from Figs. 1a and b, a big jump occurred in the sensitivities in 1985 due to a change in D3. The difference in S(*K*) of 1.99 ±0.09 mags between the 1.5 and 4m is close to the expected value of 2.13 mags. With the possible exception of S(*J*−*K*), there does not appear to be any difference greater than the uncertainty in the means for the two telescopes. Therefore, the sensitivities measured on the two telescopes will be considered together.

A final factor of interest is the state of the mirror coatings. The 4 and 1.5 meter primaries were realuminized every other year and every year, respectively. Deterioration of the coatings could cause a reduction of sensitivity in the near-IR and an increase in the thermal background emission. For the former possibility, we can set an upper limit of 10%: The standard deviation of the difference in S(*K*) from one night to the next ( except for large differences due to instrumental changes) was calculated for each telescope. Both of these values were ≤ 0.10 mags. This same quantity, calculated just for the nights on either side of an aluminization (5 times on the 4m, 8 times on the 1.5m), was not significantly different.

### 3.2 Monthly and Yearly Averages

Monthly averages for the extinction coefficients were calculated from all observations on both telescopes. Average sensitivities use data only from 1984 and earlier because of the large instrumentally caused jumps in the 1985 data. These monthly averages are given in Tables 3 and 4 and illustrated in Figures 2a and b. For each variable, the third line in the tables gives the uncertainty in the mean. Tables 5 and 6 and Figures 3a and b present yearly means for the extinction coefficients and sensitivity values, respectively. All data from both telescopes are included in calculating these means. For each variable the third line in the tables gives the uncertainty in the mean. Because of the small numbers of observations in 1986 and 1987, these years have been combined; similarly for 1991 and 1992. Since the extinction coefficients used for an observing run were average values, to calculate the yearly averages, each night in a run was considered to have the average value used for the run. Thus the values used on a long run in a year would have more weight than those determined for a short run in the same year.

## 4. TEMPORAL VARIATIONS IN THE DATA

In this section I will argue that month to month and year to year changes seen in the extinction coefficients and the sensitivity values are primarily due to seasonal and secular changes in the $H_2O$ content of the atmosphere. I will also examine the possibility that terrestrial events affect atmospheric transmission in the near-IR.

### 4.1 Monthly Averages

The monthly averages for the extinction coefficients and sensitivities are presented in Tables 3 and 4 and Figs 2a and b. Correlations that are significant at $\geq 95\%$ are in Table 7. First, note that Table 1 shows that, with the exception of $E(K-L)$ and $E(H_2O)$, the dispersion in each extinction index when *all* of the data are considered together is less than 0.02 mag. Although May through September are the wettest months at CTIO with the highest *relative* humidity, they are actually the driest in terms of column density of $H_2O$ above the observatory since these months are also the coldest. An increase in the $H_2O$ content of the atmosphere will make instrumental values of $J-K$, $H-K$, and $K-L$ blue (because atmospheric $H_2O$ has a greater effect on $L$ than on $K$ and on $K$ than on $J$ or $H$) and the $H_2O$ and CO indices red or large (because atmospheric $H_2O$ affects the 2.00 and 2.36μm filters more strongly than it does the 2.20 μm continuum filter used for both indices). An increase in atmospheric $H_2O$ will also cause instrumental $K$ magnitudes to be measured fainter, or more positive. Changes in sensitivities will be of opposite sign to the instrumental magnitudes and colors while algebraic changes in the extinction coefficients will be in the same sense as the instrumental magnitudes and colors.

The two indices with the largest overall dispersion, $E(K-L)$ and $E(H_2O)$, also show the largest seasonal variations of the kind one might expect at CTIO. Fig. 2a shows that $E(H_2O)$ is smallest during the Southern Hemisphere's winter while $E(K-L)$ is reddest. Table 7 shows that linear correlations in the expected sense, and significant at $\geq 95\%$, are indeed present between most of the monthly averaged extinction coefficients. Figure 4 illustrates some of the significant correlations between these monthly means as well as the near constancy of $E(H-K)$.

As noted above one would expect variations in the sensitivities to correlate with variations in the extinction coefficients but with opposite sign. They might be more difficult to detect, however, since sensitivity values will also be affected by even small changes in instrumental configuration. Nonetheless, Table 7 reveals that such correlations do exist. Thus, these data demonstrate the existence of seasonal variations in both overall atmospheric transmission and in the dependence of that transmission on airmass. The latter variation does not necessarily have to follow from the former if the absorption bands responsible for overall atmospheric transmission were all purely saturated ones.

## 4.2  Yearly Averages

Figs. 3a and b and Tables 5 and 6 present yearly averages for the extinction coefficients and sensitivity values.  As with the monthly averages we want to know if there are changes in these values due to changes in the Earth's atmosphere.  First, with the possible exception of E($K–L$), there are no systematic secular changes over a time span of 14 years in the extinction coefficients.  This is deduced from the fact that except for E($K–L$) the variations in the extinction coefficients over this time span can be fit by straight lines with slopes ≤0.001 ±0.001.  A fit to the E($K–L$) data, though, yields a slope of −0.003 ±0.001, a 3σ result. However, exclusion of either the first or last datum reduces this to the 2σ level, making its significance doubtful.

Are the year-to-year changes in the extinction coefficients real since several of the values differ by 3σ or more from one or both of their neighbors.  The most extreme example is the drop in E($H_2O$) for 1984 compared to 1983, a 7σ difference.  At the same time there is a 4.5σ increase in E($K–L$), and a 7σ decrease in E($K$).  The signs of these changes are consistent with that expected from a decrease in atmospheric $H_2O$ opacity.  The upper right part of Table 7 shows that 3 pairs of extinction coefficients are correlated at ≥ 95% level although none of them include E($K–L$).  Thus it would seem that some of the year to year changes in the mean extinction coefficients are due to changes in the Earth's atmosphere, most likely the $H_2O$ content, but that these changes do not show any long term pattern.

The existence of significant correlations between the extinction coefficients and the sensitivity values averaged year-by-year further support the conclusion that there are non-seasonal changes in atmospheric transparency with a time scale ≳ one year.  These correlations are listed in the middle right section of Table 7 and illustrated in Figures 5 and 6.  All of these correlations are in the sense expected given the effects of $H_2O$ on the filter bandpasses.

## 4.3 Correlations with the El Niño/Southern Oscillation (ENSO) Phenomenon

An ENSO event results from the coupled behavior of the wind pattern over the tropical Pacific Ocean and the temperature distribution of the water in the Ocean's surface layers. An excellent review of ENSO is given by Cane (1986).  Most recent work including current observations and predications can be found at the comprehensive World Wide Web site `http://www.pmel.noaa.gov/toga-tao/el-nino/home.html`. The aspects of ENSO events of importance for this investigation are the significant elevation of sea surface temperatures off the western coast of South America and the accompanying enhancement of rainfall over these coastal regions. A very strong ENSO event occurred during 1982/83.

If there were a connection between the ENSO phenomenon and atmospheric extinction and transmission in the near-IR, one would expect the $H_2O$ extinction measured at CTIO to go up at times corresponding to a strong El Niño due to the overall increase in atmospheric water content in the eastern Pacific. Two quantitative measures of the ENSO phenomenon that can be compared with the near-IR extinction and sensitivity values are: a) the Southern Oscillation Index (SOI), which measures the departure of the pressure difference between Tahiti and Darwin, Australia, from a predefined normal monthly mean; b) the NINO1+2 index which is the standardized anomaly (monthly means having again been subtracted) in the average sea surface temperature measured between latitudes 0–10S, and longitudes 80W–90W.  These are available from `ftp://nic.fb4.noaa.gov/pub/cac/cddb/indices/nino`.

Figures 7a and 7b plot 4 of the normalized extinction values for each observing run against the SOI and NINO1+2 indices. The normalization forces the mean value for each extinction coefficient to be zero and then scales them so that all have a dispersion similar to that of NINO1+2. The signs of E(*K*−*L*) were reversed to emphasize the similarity of all 4 extinction indices. The correlations between SOI, NINO1+2 and the extinction coefficients are given at the bottom of Table 7 for p values ≤ 0.10. Correlations involving E(*J*−*K*) and E(*H*−*K*) are all ≳ 0.7. This is easily understood for E(*H*−*K*) since the effects of changing the atmospheric $H_2O$ content is nearly the same in both the *H* and *K* filters (cf. Fig. 7). It is less easily understood for E(*J*−*K*) but most likely is due to a combination of significantly larger uncertainties in determining E(*J*−*K*) and a relatively small overall variation in this coefficient

The values of the SOI near −3 and of NINO1+2 between 3 and 4 occurred during the 1982/83 ENSO event and are the most extreme values recorded in a century of record keeping. Unfortunately, during a one year period that included the time when these indices were at their extrema, I had only two CTIO observing runs. Nevertheless, the correlations exhibited in Figs. 7a and 7b indicate that the ENSO phenomenon has a measurable impact on observing conditions in the near-IR via increased atmospheric $H_2O$ content. This conclusion is further supported by the fact that the *sensitivity* of the $H_2O$ index, S($H_2O$), correlates with both SOI and NINO1+2 at more than the 99% level.

Both the NINO1+2 and SOI indices also have real excursions to large negative and positive values, respectively (cf. Fig. 7, but more readily apparent in the full data sets for these indices). These excursions, indicative of atmospheric and oceanic conditions opposite those associated with an El Niño - exceptionally cold waters off the South American coast and particularly strong easterly trades across the tropical Pacific - are referred to as La Niña (Cane 1986). Figs. 7a and 7b suggest that La Niñas are associated with below normal atmospheric $H_2O$ content.

Near the start of the 1982/1983 ENSO event, the Mexican volcano El Chichón had two major eruptions (1982 March 23 and April 4). About 150 days after these eruptions the extinction at *V* increased by nearly 0.04 mags at the European Southern Observatory on La Silla, Chile (Burki et al. 1995) and took more than 1000 days to decay back down to its normal value. Unlike the near-IR extinction, though, the normal extinction at *V* , which is also observed to be seasonally variable, is due primarily to scattering and molecular diffusion (Burki et al. 1995). Burki et al. (see also Lockwood & Thompson 1986) demonstrated that the "volcanic aerosols" from the two eruptions resulted in a considerably flatter wavelength dependence in the optical but still was due primarily to scattering rather than absorption. Lockwood & Thompson (1986), based on optical extinction measurements at CTIO in the 1960s, point out that the latitude of a volcano will probably be a strong determining factor in the degree of enhancement of visible extinction at a given observing site. The present near-IR data are not adequate to address the question of whether the recent eruptions also affected extinction in the near-IR. Given that the major source of this extinction is $H_2O$, and the correlation between $H_2O$ extinction and the ENSO indices, it is doubtful that any effect due to volcanoes would be detected even with much better data sampling at the time of the eruptions. Neither Burki nor Lockwood & Thompson tested their optical data for variations due to the 1982/1983 ENSO event, but such variations seem unlikely from the graphs in their papers.

Finally, the sharp drop in S(*J*−*H*) and S(*H*−*K*) (Figure 3b and Table 6) clearly shows the change to a doublet field lens for D3 as well as a change from a Santa Barbara Research Corporation InSb detector to one by Cincinnati Electronics. These "drops" in sensitivity were accompanied by even larger reductions in detector noise so that the S/N ratio of a measurement actually *increased* for a source of given brightness and fixed integration time.

## 5. SUMMARY AND CONCLUSIONS

Since most of the atmospheric extinction in the near-IR can be attributed to absorption by $H_2O$ (see, for example, any standard atmospheric transmission curves; also Manduca & Bell 1979), one expects that if some of the observed month-to-month and year-to-year variations in the extinction coefficients and sensitivity values are intrinsic, then these variations should be correlated with one another. Table 7 shows that this is the case since many of these quantities are linearly correlated with a significance ≥95%. Most of these involve the $H_2O$ index which, because of the 2.00μm filter used in measuring it, is expected to show the greatest variation with atmospheric $H_2O$. Furthermore, the signs of the correlations are all as expected if $H_2O$ is the primary agent for varying atmospheric transmission in the near infrared. Almost all other pairings of extinction coefficients and sensitivity values on either a monthly or yearly basis produced correlations with a significance considerably less than those tabulated.

The correlations observed between the yearly means for extinction and sensitivity strongly suggests that there are real long-term changes in the near-IR transmission properties of the atmosphere. Some of these changes are correlated with events such as El Niño and La Niña which are different aspects of the more general ENSO phenomenon. The present data cannot investigate any possible effects of volcanic eruptions on near-IR extinction.

Finally, I comment on the use of mean extinction coefficients for data reduction. Depending on the accuracy desired and the difference in airmass between objects and standards, Table 1 shows that on average, mean coefficients can yield acceptable results. Table 3 and Figs. 2a and 4 show, though, that a reduction in observational uncertainty by up to several hundredths of a magnitude can be achieved by taking into account the month when the observations were made. There is no substitute, though, for determining the extinction coefficients for each observing run; Table 5 and Figs. 1a, 3a, and 5 show that even monthly means vary from year to year. Furthermore, global phenomenon such as ENSO will systematically affect the values.

Much of the observing that resulted in the data for this paper was carried out with the collaboration of Jay Elias. The able assistance of the many CTIO telescope operators and the entire CTIO TELOPS crew played an essential role in 15 years of data gathering. I thank mi hermano Oscar Saa (CTIO) for supplying me with useful information; Lonnie Thompson and Keith Henderson (Byrd Polar Research Center, OSU) for acquainting me with quantitative measures of the ENSO phenomenon; and my colleague Kris Sellgren for a critical reading of the manuscript. My research at OSU was supported in part by NSF grant AST92-18281.

# FIGURE CAPTIONS

F$_{IG}$. 1a— These extinction coefficients are the average values for each observing run based on values determined for each night within the run.   Constant shifts have been applied to some of the variables for clarity of presentation.  Typical 1 σ uncertainties are ±0.02 mag or less in all of the values.

F$_{IG}$. 1b— Sensitivities for colors are plotted for photometric nights on the CTIO 1.5 and 4 m Blanco reflectors.  A constant was added to the values for S(CO) for clarity of presentation. The sharp break observed near JD 2431000 in some of the quantities is due to changes in instrumental parameters as discussed in the text.  Uncertainties in the plotted values are comparable in size or smaller than the symbols.

F$_{IG}$. 2a— Monthly averages for the extinction coefficients from Table 3.  The inner error bars (invisible for many of the points) are the uncertainties in the mean values plotted.  The longer error bars show the variances of the means.  No data were obtained in the month of August.

F$_{IG}$. 2b— Monthly averages for the sensitivity values from Table 4. The inner error bars (invisible for many of the points) are the uncertainties in the mean values plotted.  The longer error bars show the variances of the means.  No data were obtained in the month of August.

F$_{IG}$. 3a— Extinction coefficients averaged for the years indicated with uncertainties in the mean values.  Data are from Table 5.

F$_{IG}$. 3b— Sensitivities averaged for the years indicated with uncertainties in the mean values.  Data are from Table 6.  Individual values may be seen in Fig. 1b.

F$_{IG}$. 4— Table 7 shows that significant correlations exist between the monthly means of $E(H_2O)$ and the other extinction coefficients except for $E(H-K)$.  This figure illustrates these correlations between the monthly mean values and also shows the simple, unweighted, least squares regression line for each.  Error bars are the uncertainties in the mean values as given in Table 3.  Note the shift between the left and right hand axes.

F$_{IG}$. 5— Similar to Figure 4 except correlations between the yearly means are illustrated. Again note the shift between the left and right hand axes.

F$_{IG}$. 6— Similar to Figure 5.  Note the difference scales for the bottom-right and top-left.

F$_{IG}$. 7a,b— These figures display average extinction values for each run normalized to zero mean and with dispersions close to that of the NINO1+2 index as functions of both the NINO1+2 and SOI indices.  The NINO1+2 index is an indicator of  how high above normal for any given month the sea surface temperature is between 0–10S and 80–90W.  The SOI index is the departure from normal of the difference in barometric pressure between Tahiti and Darwin, Australia.  The time period covered in 1978 to 1992.  The dashed lines are the regression solutions for each of the 4 extinction coefficients.  For NINO1+2 as the independent variable, the slopes, in the order indicated by the legend, are 0.327, 0.234, 0.355, and 0.238.  For SOI as the independent variable, the slopes, in the same order, are -0.359, –0.601, –0.358, and –0.326.

Table 1

A Comparison of Results from the 4-m and 1.5-m: Extinction Coefficients

| | Extinction Coefficients (all data) | | | | | | | |
| | 1.5 m | | | 4 m | | | Combined | |
| | # nights | mean | σ | # nights | mean | σ | mean | σ |
|---|---|---|---|---|---|---|---|---|
| E($J$–$K$) | 70 | 0.011 | 0.018 | 125 | 0.018 | 0.014 | 0.015 | 0.016 |
| E($H$–$K$) | 70 | -0.029 | 0.011 | 125 | -0.030 | 0.010 | -0.030 | 0.010 |
| E($K$) | 70 | 0.081 | 0.016 | 125 | 0.087 | 0.019 | 0.085 | 0.018 |
| E($K$–$L$) | 62 | -0.061 | 0.021 | 105 | -0.071 | 0.026 | -0.067 | 0.025 |
| E($H_2O$) | 67 | 0.236 | 0.030 | 118 | 0.242 | 0.028 | 0.240 | 0.029 |
| E(CO) | 67 | 0.072 | 0.013 | 118 | 0.070 | 0.011 | 0.071 | 0.012 |

Table 2

A Comparison of Results from the 4-m and 1.5-m: Sensitivities

| | Sensitivities (1978 - 1984) | | | | | |
| | 1.5 m | | | 4 m | | |
| | # nights | mean | σ | # nights | mean | σ |
|---|---|---|---|---|---|---|
| S($J$–$K$) | 63 | 0.64 | 0.12 | 100 | 0.58 | 0.12 |
| S($H$–$K$) | 63 | 0.47 | 0.04 | 100 | 0.47 | 0.05 |
| S($K$) | 63 | 16.91 | 0.57 | 100 | 18.90 | 0.50 |
| S($K$–$L$) | 58 | 1.45 | 0.10 | 82 | 1.47 | 0.09 |
| S($H_2O$) | 61 | -0.58 | 0.08 | 96 | -0.56 | 0.08 |
| S(CO) | 59 | -0.60 | 0.03 | 96 | -0.59 | 0.03 |

TABLE 3
Monthly Averages: Extinction Coefficients

| | | Extinction Coefficients (all data) | | | | | | | | | | |
|---|---|---|---|---|---|---|---|---|---|---|---|---|
| | | Jan | Feb | Mar | Apr | May | Jun | Jul | Sep | Oct | Nov | Dec |
| E(J–K) | # nights | 23 | 18 | 17 | 12 | 33 | 9 | 5 | 3 | 13 | 24 | 38 |
| | mean | 0.010 | 0.008 | 0.002 | 0.007 | 0.022 | 0.013 | 0.020 | 0.020 | 0.021 | 0.025 | 0.016 |
| | $\sigma/\sqrt{N}$ | 0.002 | 0.002 | 0.004 | 0.005 | 0.003 | 0.005 | 0.012 | 0.000 | 0.006 | 0.002 | 0.003 |
| E(H–K) | # nights | 23 | 18 | 17 | 12 | 33 | 9 | 5 | 3 | 13 | 24 | 38 |
| | mean | -0.034 | -0.028 | -0.036 | -0.029 | -0.027 | -0.028 | -0.028 | -0.030 | -0.022 | -0.030 | -0.029 |
| | $\sigma/\sqrt{N}$ | 0.002 | 0.002 | 0.003 | 0.003 | 0.002 | 0.001 | 0.007 | 0.000 | 0.002 | 0.002 | 0.002 |
| E(K) | # nights | 23 | 18 | 17 | 12 | 33 | 9 | 5 | 3 | 13 | 24 | 38 |
| | mean | 0.087 | 0.088 | 0.097 | 0.088 | 0.077 | 0.063 | 0.078 | 0.070 | 0.098 | 0.082 | 0.087 |
| | $\sigma/\sqrt{N}$ | 0.002 | 0.004 | 0.005 | 0.003 | 0.002 | 0.002 | 0.005 | 0.000 | 0.009 | 0.003 | 0.003 |
| E(K–L) | # nights | 23 | 17 | 13 | 11 | 27 | 9 | 5 | 3 | 11 | 16 | 32 |
| | mean | -0.069 | -0.066 | -0.074 | -0.079 | -0.076 | -0.050 | -0.036 | -0.030 | -0.064 | -0.061 | -0.071 |
| | $\sigma/\sqrt{N}$ | 0.005 | 0.005 | 0.010 | 0.003 | 0.003 | 0.010 | 0.002 | 0.000 | 0.006 | 0.004 | 0.005 |
| E(H$_2$O) | # nights | 23 | 16 | 17 | 12 | 30 | 9 | 5 | 3 | 13 | 20 | 37 |
| | mean | 0.246 | 0.269 | 0.257 | 0.260 | 0.222 | 0.229 | 0.212 | 0.210 | 0.248 | 0.214 | 0.243 |
| | $\sigma/\sqrt{N}$ | 0.003 | 0.008 | 0.004 | 0.006 | 0.005 | 0.006 | 0.007 | 0.000 | 0.008 | 0.005 | 0.004 |
| E(CO) | # nights | 21 | 16 | 17 | 12 | 32 | 9 | 5 | 3 | 13 | 20 | 37 |
| | mean | 0.075 | 0.079 | 0.074 | 0.075 | 0.067 | 0.064 | 0.076 | 0.060 | 0.071 | 0.058 | 0.073 |
| | $\sigma/\sqrt{N}$ | 0.001 | 0.004 | 0.001 | 0.002 | 0.002 | 0.003 | 0.010 | 0.000 | 0.003 | 0.002 | 0.002 |

TABLE 4
Monthly Averages: Sensitivities

| Sensitivity Values (1978 - 1984) | | | | | | | | | | | |
|---|---|---|---|---|---|---|---|---|---|---|---|
| | Jan | Feb | Mar | Apr | May | Jun | Jul | Sep | Oct | Nov | Dec |
| S($J$–$K$) # nights | 23 | 18 | 15 | 10 | 21 | 6 | 5 | 3 | 13 | 13 | 36 |
| mean | 0.62 | 0.50 | 0.60 | 0.70 | 0.65 | 0.50 | 0.52 | 0.78 | 0.59 | 0.68 | 0.59 |
| $\sigma/\sqrt{N}$ | 0.02 | 0.03 | 0.04 | 0.02 | 0.03 | 0.01 | 0.02 | 0.01 | 0.04 | 0.01 | 0.02 |
| S($H$–$K$) # nights | 23 | 18 | 15 | 10 | 21 | 6 | 5 | 3 | 13 | 13 | 36 |
| mean | 0.47 | 0.45 | 0.45 | 0.50 | 0.48 | 0.48 | 0.47 | 0.46 | 0.46 | 0.48 | 0.48 |
| $\sigma/\sqrt{N}$ | 0.01 | 0.00 | 0.01 | 0.02 | 0.01 | 0.00 | 0.00 | 0.01 | 0.02 | 0.02 | 0.01 |
| S($K$–$L$) # nights | 23 | 17 | 11 | 9 | 17 | 6 | 5 | 3 | 10 | 9 | 30 |
| mean | 1.50 | 1.53 | 1.49 | 1.54 | 1.40 | 1.47 | 1.44 | 1.53 | 1.42 | 1.34 | 1.45 |
| $\sigma/\sqrt{N}$ | 0.01 | 0.02 | 0.02 | 0.08 | 0.02 | 0.01 | 0.01 | 0.02 | 0.04 | 0.01 | 0.02 |
| S($H_2O$) # nights | 23 | 16 | 15 | 9 | 20 | 6 | 5 | 3 | 13 | 13 | 34 |
| mean | -0.60 | -0.65 | -0.61 | -0.59 | -0.52 | -0.50 | -0.47 | -0.52 | -0.56 | -0.50 | -0.58 |
| $\sigma/\sqrt{N}$ | 0.01 | 0.03 | 0.01 | 0.01 | 0.02 | 0.01 | 0.01 | 0.02 | 0.02 | 0.01 | 0.01 |
| S(CO) # nights | 21 | 16 | 15 | 9 | 20 | 6 | 5 | 3 | 13 | 13 | 34 |
| mean | -0.60 | -0.61 | -0.59 | -0.61 | -0.59 | -0.56 | -0.56 | -0.59 | -0.58 | -0.60 | -0.60 |
| $\sigma/\sqrt{N}$ | 0.01 | 0.01 | 0.00 | 0.01 | 0.00 | 0.00 | 0.00 | 0.00 | 0.01 | 0.01 | 0.01 |

TABLE 5

Yearly Averages:  Extinction Coefficients

| | | Extinction Coefficients | | | | | | | | | |
|---|---|---|---|---|---|---|---|---|---|---|---|
| | | 1978 | 1979 | 1980 | 1981 | 1982 | 1983 | 1984 | 1985 | 1986/7 | 1991/2 |
| E($J$–$K$) | # nights | 12 | 35 | 27 | 32 | 19 | 14 | 22 | 18 | 10 | 6 |
| | mean | 0.025 | 0.003 | 0.021 | 0.015 | 0.003 | 0.025 | 0.028 | 0.014 | 0.006 | 0.033 |
| | $\sigma / \sqrt{N}$ | 0.002 | 0.003 | 0.002 | 0.003 | 0.001 | 0.003 | 0.001 | 0.004 | 0.003 | 0.002 |
| E($H$–$K$) | # nights | 12 | 35 | 27 | 32 | 19 | 14 | 22 | 18 | 10 | 6 |
| | mean | -0.025 | -0.037 | -0.027 | -0.029 | -0.034 | -0.024 | -0.019 | -0.033 | -0.040 | -0.017 |
| | $\sigma / \sqrt{N}$ | 0.002 | 0.002 | 0.002 | 0.002 | 0.001 | 0.001 | 0.001 | 0.002 | 0.003 | 0.002 |
| E($K$) | # nights | 12 | 35 | 27 | 32 | 19 | 14 | 22 | 18 | 10 | 6 |
| | mean | 0.070 | 0.099 | 0.091 | 0.077 | 0.087 | 0.091 | 0.068 | 0.082 | 0.090 | 0.083 |
| | $\sigma / \sqrt{N}$ | 0.000 | 0.004 | 0.001 | 0.002 | 0.003 | 0.001 | 0.003 | 0.004 | 0.009 | 0.004 |
| E($K$–$L$) | # nights | 12 | 23 | 23 | 29 | 18 | 14 | 22 | 14 | 9 | 3 |
| | mean | -0.030 | -0.059 | -0.068 | -0.069 | -0.066 | -0.093 | -0.073 | -0.066 | -0.077 | -0.093 |
| | $\sigma / \sqrt{N}$ | 0.000 | 0.004 | 0.003 | 0.006 | 0.007 | 0.004 | 0.002 | 0.001 | 0.004 | 0.007 |
| E($H_2O$) | # nights | 12 | 35 | 25 | 31 | 19 | 14 | 21 | 14 | 10 | 4 |
| | mean | 0.225 | 0.238 | 0.250 | 0.252 | 0.252 | 0.257 | 0.210 | 0.221 | 0.240 | 0.240 |
| | $\sigma / \sqrt{N}$ | 0.005 | 0.005 | 0.004 | 0.006 | 0.004 | 0.005 | 0.004 | 0.006 | 0.010 | 0.023 |
| E(CO) | # nights | 12 | 35 | 25 | 31 | 17 | 14 | 21 | 14 | 10 | 6 |
| | mean | 0.060 | 0.067 | 0.081 | 0.075 | 0.075 | 0.073 | 0.061 | 0.060 | 0.081 | 0.070 |
| | $\sigma / \sqrt{N}$ | 0.000 | 0.002 | 0.001 | 0.002 | 0.003 | 0.002 | 0.001 | 0.000 | 0.003 | 0.000 |

TABLE 6
Yearly Averages: Sensitivities

| | | Sensitivity Values | | | | | | | | | |
|---|---|---|---|---|---|---|---|---|---|---|---|
| | | 1978 | 1979 | 1980 | 1981 | 1982 | 1983 | 1984 | 1985 | 1986/7 | 1991/2 |
| S($J$–$K$) | # nights | 12 | 35 | 27 | 32 | 19 | 16 | 22 | 18 | 10 | 6 |
| | mean | 0.67 | 0.73 | 0.62 | 0.44 | 0.48 | 0.63 | 0.69 | -0.10 | -0.07 | -0.02 |
| | $\sigma/\sqrt{N}$ | 0.02 | 0.01 | 0.02 | 0.01 | 0.01 | 0.02 | 0.01 | 0.01 | 0.01 | 0.02 |
| S($H$–$K$) | # nights | 12 | 35 | 27 | 32 | 19 | 16 | 22 | 18 | 10 | 6 |
| | mean | 0.39 | 0.45 | 0.47 | 0.46 | 0.47 | 0.53 | 0.54 | 0.12 | 0.16 | 0.16 |
| | $\sigma/\sqrt{N}$ | 0.00 | 0.00 | 0.00 | 0.00 | 0.00 | 0.01 | 0.00 | 0.00 | 0.00 | 0.01 |
| S($K$–$L$) | # nights | 12 | 23 | 23 | 29 | 18 | 13 | 22 | 14 | 9 | 3 |
| | mean | 1.42 | 1.47 | 1.47 | 1.53 | 1.54 | 1.43 | 1.35 | 1.27 | 1.16 | 1.40 |
| | $\sigma/\sqrt{N}$ | 0.02 | 0.04 | 0.01 | 0.01 | 0.01 | 0.01 | 0.01 | 0.01 | 0.20 | 0.01 |
| S($H_2O$) | # nights | 12 | 35 | 25 | 31 | 19 | 14 | 21 | 14 | 10 | 4 |
| | mean | -0.53 | -0.57 | -0.59 | -0.58 | -0.62 | -0.57 | -0.50 | -0.57 | -0.62 | -0.56 |
| | $\sigma/\sqrt{N}$ | 0.01 | 0.01 | 0.01 | 0.02 | 0.02 | 0.01 | 0.02 | 0.01 | 0.03 | 0.04 |
| S(CO) | # nights | 12 | 35 | 25 | 31 | 17 | 14 | 21 | 14 | 10 | 6 |
| | mean | -0.56 | -0.59 | -0.58 | -0.58 | -0.60 | -0.62 | -0.64 | -0.57 | -0.59 | -0.58 |
| | $\sigma/\sqrt{N}$ | 0.01 | 0.00 | 0.00 | 0.00 | 0.01 | 0.01 | 0.01 | 0.00 | 0.01 | 0.01 |

Table 7
Tests for Correlations

| Monthly Means | | Yearly Means (1978-1992) | |
|---|---|---|---|
| Pair of Variables | p-levels | Pair of Variables | p-levels |
| E($H_2O$) & E($J$–$K$) | -0.00 | E($J$–$K$) & E($H$–$K$) | +0.00 |
| E($H_2O$) & E($K$) | +0.02 | E($H_2O$) & E(K) | +0.05 |
| E($H_2O$) & E($K$–$L$) | -0.02 | E($H_2O$) & E(CO) | +0.01 |
| E($H_2O$) & E(CO) | +0.01 | | |
| E($J$–$K$) & E(CO) | -0.03 | | |
| E($K$–$L$) & E($K$) | -0.04 | Yearly Means (1978-1984) | |
| | | Pair of Variables | p-levels |
| S($H_2O$) & S(CO) | +0.04 | | |
| | | S($H_2O$) & S($K$–$L$) | -0.00 |
| S($H_2O$) & E($J$–$K$) | +0.01 | S(CO) & S($H$–$K$) | -0.00 |
| S($H_2O$) & E($K$) | -0.02 | | |
| S($H_2O$) & E($K$–$L$) | +0.04 | S($H$–$K$) & E($K$–$L$) | -0.01 |
| S($H_2O$) & E($H_2O$) | -0.00 | S($H_2O$) & E($H_2O$) | -0.01 |
| S($H_2O$) & E(CO) | -0.03 | S($H_2O$) & E(CO) | -0.02 |
| S($K$–$L$) & E($J$–$K$) | -0.01 | S($K$–$L$) & E($H_2O$) | +0.05 |

| With ENSO Indices | | | |
|---|---|---|---|
| Pair of Variables | p-levels | Pair of Variables | p-levels |
| SOI & E($K$) | 0.029 | NINO1+2 & E($K$) | 0.015 |
| SOI & E($H_2O$) | 0.000 | NINO1+2 & E($H_2O$) | 0.010 |
| SOI & E($K$–$L$) | 0.032 | NINO1+2 & E($K$–$L$) | 0.10 |
| SOI & E(CO) | 0.05 | NINO1+2 & E(CO) | 0.09 |